\documentclass{llncs}
\usepackage{makeidx}  % allows for indexgeneration

\usepackage[dvips]{graphicx}
\usepackage{timet}
\usepackage{amsmath}
\usepackage{url}
\usepackage{algorithm}
\usepackage{algorithmic}
\usepackage{multirow}

\makeindex
\begin{document}
\frontmatter          % for the preliminaries
\setcounter{page}{1}
\pagestyle{headings}  % switches on printing of running heads
\addtocmark{Hamiltonian Mechanics} % additional mark in the TOC
%
% \title{Low-Order Implicit Unstructured Finite-Element Multiple Simulation Enhanced by Dense Computation using OpenACC}
\title{Implicit Low-Order Unstructured Finite-Element Multiple Simulation Enhanced by Dense Computation using OpenACC}
%
%\titlerunning{Hamiltonian Mechanics}  % abbreviated title (for running head)
%                                     also used for the TOC unless
%                                     \toctitle is used
%
\author{Takuma Yamaguchi\inst{1} \and Kohei Fujita\inst{1,2} \and Tsuyoshi Ichimura\inst{1,2} \and Muneo Hori\inst{1,2} \and Maddegedara Lalith\inst{1,2} \and Kengo Nakajima\inst{3}}
%
%\authorrunning{Ivar Ekeland et al.}   % abbreviated author list (for running head)
%
%%%% list of authors for the TOC (use if author list has to be modified)
\tocauthor{Takuma Yamaguchi, Kohei Fujita, Tsuyoshi Ichimura, Munoe Hori, Mageddedara Lalith, and Kengo Nakajima}

% use the command \index{<name>} for index entries
%
\institute{Earthquake Research Institute and Department of Civil Engineering,\\ The University of Tokyo, 1-1-1 Yayoi, Bunkyo, Tokyo, Japan,\\
\email{\{yamaguchi, fujita, ichimura, hori, lalith\}@eri.u-tokyo.ac.jp}
\and
Advanced Institute for Computational Science, RIKEN
\and
%Supercomputing Division 
Information Technology Center, The University of Tokyo, \\2-11-16 Yayoi, Bunkyo, Tokyo, Japan,\\
\email{nakajima@cc.u-tokyo.ac.jp}}

\maketitle              % typeset the title of the contribution

\newcommand{\SOLVER}{GAMERA+AMG}

\section*{Abstract}

In this paper, we develop a low-order three-dimensional finite-element solver for fast multiple-case crust deformation computation on GPU-based systems.
Based on a high-performance solver designed for massively parallel CPU-based systems, we modify the algorithm to reduce random data access, and then insert OpenACC directives.
%それぞれのアーキテクチャに合わせたアルゴリズムを開発することによって，GPUの性能をより引き出すことに成功した．
By developing algorithm appropriate for each computer architecture, we enable to exhibit higher performance.
The developed solver on ten Reedbush-H nodes (20 P100 GPUs) attained speedup of 14.2 times from the original solver on 20 K computer nodes. 
On the newest Volta generation V100 GPUs, the solver attained a further 2.52 times speedup with respect to P100 GPUs.
As a demonstrative example, we computed 368 cases of crustal deformation analyses of northeast Japan with 400 million degrees of freedom.
The total procedure of algorithm modification and porting implementation took only two weeks; we can see that high performance improvement was achieved with low development cost.
With the developed solver, we can expect improvement in reliability of crust-deformation analyses by many-case analyses on a wide range of GPU-based systems.

\section{Introduction}
\label{sct1}
% 地震災害の一連のプロセスへの理解を深め，予測の信頼性を高めるために，我々は地震災害の高精度・高分解能シミュレーションを開発してきた．
% We have been developing a high accurate and high resolution earthquake disaster simulator for gaining knowledge on earthquake disaster processes and improve estimation accuracy of future earthquakes.
Simulations reflecting the physical phenomena of earthquake disasters are useful for gaining knowledge on earthquake disaster processes and improve estimation accuracy for future earthquakes.
% 対象となる領域は広範囲かつ複雑な形状を有するため，陰解法・低次要素非構造格子を用いた大規模三次元有限要素解析が適している．
As the target domain of earthquake disaster simulations is heterogeneous and involves complex geometry, large-scale implicit three-dimensional (3D) finite-element analysis using low-order unstructured elements is suitable.
% この解析を実用につなげるためには解析の高速化が必要となるが，低次要素を用いた解析のため一般的には高い性能を出すことは難しいと考えられる．
% 解析の計算コストの大部分は連立一次方程式の求解となる．
In such simulations, most of the computing cost is spent in solving a large system of linear equations.
% 我々はこの計算時間を削減するために高速なソルバーを開発しており，CPUベースの大規模計算機環境である京コンピューター~\cite{refK}を対象として開発されたGAMERA, GOJIRAについてはそれぞれSC14, SC15にてGordon Bell finalistに選出されている．~\cite{gamera},~\cite{gojira}
Thus, we have been developing fast solver algorithms for CPU-based systems; our solvers GAMERA and GOJIRA running on the massively parallel CPU-based K computer system~\cite{refK} were nominated as Gordon Bell Prize finalists in SC14 and SC15 \cite{gamera}, \cite{gojira}.
% WACCPD2016では，我々は高効率な計算と計算機環境の拡張を目的としてこのGOJIRAをGPUへ移植し，OpenACC~\cite{openacc}を用いたGPU計算の有用性を確認した~\cite{waccpd2016}．
Furthermore, we ported this solver to GPU environments using OpenACC \cite{openacc} , which was presented at the Workshop on Accelerator Programming Using Directives (WACCPD) 2016 \cite{fujita2016acceleration}.
This enabled further acceleration and use on wider computing environments with low additional development cost.

% 地震災害の被害軽減のための有効なアプローチの一つとして，現在地震発生の時間・位置・大きさを予測することが検討されている．
%Targeting further improvement in earthquake mitigation, prediction of the time, position, and magnitude of an earthquake is now being considered.
Although very challenging, forecasting of the time, position, and magnitude of an earthquake is one of the major goals in earthquake science and disaster mitigation.
One promising means of such forecasting could be physics-based forecasting that uses GPS observation data of crust-deformation and many-case analyses to estimate plate boundary states via inverse analyses.
% これはGPS等による地殻変動観測網を用いた逆解析によってプレート境界の状態を推定し，地震発生の可能性を予測するものである．
%Here, GPS observation data of crust deformation is used to estimate plate boundary states using inverse analyses, and these information are used to estimate the potential of future earthquakes.
% この逆解析には与えられた滑り量に対する地表面応答が多数必要となる．
In these analyses, many cases of 3D finite-element simulations for each slip distribution are required.
% そのため大規模三次元有限要素解析による地殻変動計算を多数回繰り返して行う必要があり，シミュレーションに要求される計算量はさらに増大する．
Thus, the required computing cost increases significantly when compared with single case finite-element simulations.
% この計算量に対処するためにGAMERAをもとにしたアルゴリズムの拡張が進められており，地殻変動計算へと特化したソルバーとして，CPUベースのソルバー\SOLVER が開発されている~\cite{pasc}．
To deal with the increased computational costs, an algorithm that is specialized for crust-deformation analysis has been developed based on GAMERA \cite{pasc}.
% これはGAMERAの2.18倍の性能となることが確認されている．
This solver enabled 2.18 times speedup on crust-deformation problems on the K computer when compared with GAMERA.
% このソルバーをOpenACCを用いてポートすることにより，少ない開発コストによって地殻変動計算のさらなる高速化が実現すると期待できる．
Porting this solver to GPU-based systems using OpenACC can be expected to result in further speedup of crust-deformation analysis with small development cost.
% 一方でGPUは一般にランダムアクセスを伴う計算ではメモリアクセスのレイテンシが顕著に表れることが知られており，通常の有限要素解析ではメモリバンド幅律速となる．
On the other hand, GPUs are known to involve large memory access latencies for random accesses, and in addition, standard finite-element applications tend to be memory bandwidth bound.
% そのためCPUのコードをそのまま移植するのみでは，GPUの高い演算性能を十分に活用することは難しい．
Thus, simple porting of the CPU code is not sufficient to utilize the high computing capability of GPUs.
% そこで我々はGPUの弱点であるランダムメモリアクセスが軽減されるように解析を組み替える．
Thus, we change the computational order of calculation such that random memory access can be reduced when porting the solver to GPUs.
% すなわち本研究では\SOLVER をOpenACCでポートすると共にGPU用にアルゴリズムを変更することで更なる高速化を実現させる．
% In particular, we port this solver using OpenACC together with algorithm modification.
% このアプローチが最新のアーキテクチャに対しても有効であることを示すために，我々はPascal GPU~\cite{pascal}だけでなくVolta GPU~\cite{volta}を用いて性能の評価を行う．
To show the effectiveness of the solver on the newest architecture, we measure performance on the Pascal \cite{pascal} and Volta \cite{volta} generation GPUs.
% Volta GPUは現行のアーキテクチャであるPascal GPUと比較するとさらにFLOPS比メモリバンド幅が減少しているため，本アプローチがより大きく影響することが示される(とよい)
As the Volta GPUs have less memory throughput per floating-point computation capability, we expect higher effectiveness of our method on these GPUs.
% また適用例の計算として，東北沖地震時滑り量分布の推定を行い，実際の問題に対しても高速に計算が行えることを示す．
As a demonstration of the developed method, we estimate slip distribution during the 2011 Tohoku-oki earthquake.

% 以下に本論文の構成を説明する．
The remainder of the paper is as follows.
% 2章では京コンピューター用に開発された，高速な有限要素ソルバー\SOLVER について説明する．
Section \ref{sct2} summarizes the CPU-based finite-element solver developed in \cite{pasc} for the K computer.
% 3章ではOpenACCを導入し，解析の順番を組み替えることによって，さらなる高速化を実現する．
Section \ref{sct3} explains the algorithm changes and the use of OpenACC for acceleration of the solver on GPUs.
% 4章では開発したソルバーに対して最新のアーキテクチャであるVolta GPUと，Pascal GPUとを用いた性能の比較を行う．
Section \ref{sct4} explains the performance of the developed solver on the newest Volta GPUs and recent Pascal GPUs.
% 5章では開発したソルバーを用いて，実問題の求解を行う．(東北沖地震時滑り量分布の推定)
Section \ref{sct5} shows an application example using the developed solver on a Tohoku-oki earthquake problem.
% 6章はまとめとなる．
Section \ref{sct6} summarizes the paper.

\section{Finite-Element Earthquake Simulation Designed for the K Computer}
\label{sct2}
%\paragraph{{\bf Target Equation}}
% 震源とその地表面変位応答をモデル化する場合，現象としての時間スケールは数時間から数日と短いため，解析対象であるリソスフェア・アセノスフェアはいずれとも線形弾性体とみなせる．
As the time scale of crust-deformation due to faulting is a few hours to a few days, we can regard the target crust (including the lithosphere and the asthenosphere) as a linear elastic solid.
% 地殻構造の3次元有限要素モデル内の断層面上の節点に強制変位を与えた時の，地表面に存在する節点の変位を求める静弾性解析としてモデル化を行う．
Here we analyze the static elastic response at the surface given a slip distribution at the fault plane.
% すなわち地殻構造は以下に示される線形弾性体の基礎方程式に従うとみなす，
This follows the governing equations below:
\begin{subequations}
\begin{equation}
  \epsilon_{ij} = \cfrac{1}{2} \left( \cfrac{\partial u_i}{\partial x_j}+\cfrac{\partial u_j}{\partial x_i} \right), \\
\end{equation}
\begin{equation}
  \sigma_{ij}=C_{ijkl}\epsilon_{kl}, \\
\end{equation}  
\begin{equation}
  \cfrac{\partial \sigma_{ij}}{\partial x_j}= 0 .
\end{equation}
\end{subequations}
% ただし $\epsilon_{ij}$は線形ひずみ，$u_i$は変位， $C_{ijkl}$は弾性係数テンソル，$\sigma_{ij}$は応力である．
Here, $\epsilon_{ij}$ is the elastic strain, $u_i$ is the displacement, $C_{ijkl}$ is the elastic coefficient tensor, and $\sigma_{ij}$ is the stress.
% これらの基礎方程式に境界条件と，設定したすべり面からsplit node technique~\cite{melosh1981simple}を用いて求められる節点力を加えることで，地殻構造モデルの挙動が決まる．
% 基礎方程式を2次の補間関数を用いて離散化すれば，モデルの挙動を求める問題は以下の連立一次方程式を解く事に帰着する．
By discretizing the governing equation using second ordered tetrahedral elements, we obtain
\begin{equation}
{\bf K}{\bf u} = {\bf f}, \label{eqFGE}
\end{equation}
%ただし ${\bf K}$, ${\bf u}$, and ${\bf f}$ はそれぞれ全体剛性マトリクス，変位ベクトル，外力ベクトルである．\\
where ${\bf K}$, ${\bf u}$, and ${\bf f}$ are the global stiffness matrix, displacement vector, and force vector, respectively.
We can compute the response of the crust structure model by setting the boundary condition based on the given slip at the fault using the split-node technique \cite{melosh1981simple}.

% シミュレーションを行う上ではこの連立一次方程式を高速に解くことが必要となる．
Most of the cost in finite-element analysis involves solving Eq.~(\ref{eqFGE}).
Since the finite-element model of crust structure can have as many as billions of degrees of freedom, a fast and scalable solver capable of utilizing large supercomputer systems is required.
% 京コンピューターのような超大規模計算機環境にて高い性能を出すソルバーを開発するためには，収束性の高い前処理の導入，footprintの削減および並列性の向上が必要となる．
Thus, we have designed an algorithm that attains high convergence of an iterative solver with low computation and communication cost with a small memory footprint for use on the K computer system \cite{pasc}. 
% これらを踏まえて開発されたのが，\SOLVER となる．
% GAMERA+AMG has been developed based on these requirements.
% ソルバーの構成はアルゴリズム~\ref{algo}に示される．
The algorithm of this CPU-based solver is shown in Algorithm~\ref{algo}.
Below, we summarize the key concepts used in the solver.

\begin{algorithm}
% \DontPrintSemicolon
\caption{The iterative solver is calculated to obtain a converged solution of ${\bf Ku=f}$ using an initial solution, ${\bf u}$, with a threshold of $\|{\bf Ku} - {\bf f}\|^{2}/\|{\bf f}\|^{2} \leq \epsilon$. The input variables are: ${\bf u}, {\bf f}, {\bf K}, \overline{\bf K}, \overline{\bf K}_{1}, \overline{\bf A}_{2}, \overline{\bf P}_{1-2}, \epsilon, \overline{\epsilon}_{0-2}$, and $N_{0-2}$. The other variables are temporal. $\overline{\bf P}_{1-2}$ are mapping matrices from the coarser model to the finer model. {\it diag}[\ \ \ ], $\epsilon$, and $N$ indicate a 3$\times$3 block Jacobi of [\ \ \ ], tolerance for relative error, and maximum number of iterations, respectively. $(\bar{\ \ \ })$ represents the single-precision variables, while the others represent the double-precision variables.}
\label{algo}
\centering

\begin{minipage}{0.585\textwidth}
\textbf{(a) Outer loop}\; \\
\begin{algorithmic}[1]
\STATE set $\overline{\bf M}_{0} \Leftarrow diag[\overline{\bf K}]$\;
\STATE set $\overline{\bf M}_{1} \Leftarrow diag[\overline{\bf K}_{1}]$\;
\STATE set $\overline{\bf M}_{2} \Leftarrow diag[\overline{\bf A}_{2}]$\;
\STATE {${\bf r} \Leftarrow \sum_{i}{\bf K}_{e}^{i}{\bf u}_{e}^{i}$}\;
\STATE ${\bf r} \Leftarrow {\bf f-r}$\;
\STATE $\beta \Leftarrow 0$\;
\STATE $i \Leftarrow 1$\; 
\WHILE{$\|{\bf r}\|^{2}/\|{\bf f}\|^{2} > \epsilon$}
\STATE {$\overline{\bf r} \Leftarrow {\bf r}$}\;
\STATE {$\overline{\bf u} \Leftarrow \overline{\bf M}_{0}^{-1}\overline{\bf r}$}\;
\STATE {$\overline{\bf r}_{1} \Leftarrow \overline{\bf P}^{T}_{1}\overline{\bf r}$},\ \ {$\overline{\bf u}_{1} \Leftarrow \overline{\bf P}^{T}_{1}\overline{\bf u}$}\;
\STATE {$\overline{\bf r}_{2} \Leftarrow \overline{\bf P}^{T}_{2}\overline{\bf r}_{1}$},\ \ {$\overline{\bf u}_{2} \Leftarrow \overline{\bf P}^{T}_{2}\overline{\bf u}_{1}$}\; 
\STATE solve $\overline{\bf u}_{2}=\overline{\bf A}_{2}^{-1}\overline{\bf r}_{2}$ using {\bf (b)} with $\overline{\epsilon}_{2}$ and $N_{2}$ \\ *{\it inner loop level 2} \; 
\STATE {$\overline{\bf u}_{1} \Leftarrow \overline{\bf P}_{2}\overline{\bf u}_{2}$}\; 
\STATE solve $\overline{\bf u}_{1}=\overline{\bf K}_{1}^{-1}\overline{\bf r}_{1}$ using {\bf (b)} with $\overline{\epsilon}_{1}$ and $N_{1}$\\ *{\it inner loop level 1} \; 
\STATE {$\overline{\bf u} \Leftarrow \overline{\bf P}_{1}\overline{\bf u}_{1}$}\; 
\STATE solve $\overline{\bf u}=\overline{\bf K}^{-1}\overline{\bf r}$ using {\bf (b)} with $\overline{\epsilon}_{0}$ and $N_{0}$\\ *{\it inner loop level 0} \;
\STATE {${\bf u} \Leftarrow \overline{\bf u}$}\;
\IF {$i>1$}
\STATE {$\gamma \Leftarrow ({\bf z}, {\bf q})$}\;
\STATE $\beta \Leftarrow \gamma/\rho$\;
\ENDIF
\STATE {${\bf p} \Leftarrow {\bf z}+\beta {\bf p}$}\;
\STATE {${\bf q} \Leftarrow \sum_{i}{\bf K}_{e}^{i}{\bf p}_{e}^{i}$}\;
\STATE $\rho \Leftarrow ({\bf z}, {\bf r})$\;
\STATE $\gamma \Leftarrow ({\bf p}, {\bf q})$\;
\STATE $\alpha \Leftarrow \rho/\gamma$\;
\STATE ${\bf r} \Leftarrow {\bf r} - \alpha {\bf q}$\;
\STATE ${\bf u} \Leftarrow {\bf u} + \alpha {\bf p}$\;
\STATE $i \Leftarrow i+1$\;
\ENDWHILE
\end{algorithmic}
%\vspace{5mm}
\end{minipage}
\begin{minipage}{0.315\textwidth}
\textbf{(b) Inner loop}\;\\
\begin{algorithmic}[1]
\STATE {$\overline{\bf e} \Leftarrow \overline{\bf K}({\rm or} \overline{\bf A}) \overline{\bf u}$}\;
\STATE {$\overline{\bf e} \Leftarrow \overline{\bf r}-\overline{\bf e}$}\;
\STATE {$\overline{\beta} \Leftarrow 0$}\;
\STATE {$i \Leftarrow 1$}\;
\WHILE {$\|\overline{\bf e}\|^{2}/\|\overline{\bf r}\|^{2}>\overline{\epsilon}$ \\ \hspace{2.3em} \  {\rm and} \  $i<N$ }
\STATE {$\overline{\bf z} \Leftarrow \overline{\bf M}^{-1}\overline{\bf e}$}\;
\STATE {$\overline{\rho}_{a} \Leftarrow (\overline{\bf z}, \overline{\bf e})$}\;
\IF {$i>1$}
\STATE $\overline{\beta} \Leftarrow \overline{\rho}_{a}/\overline{\rho}_{b}$\;
\ENDIF
\STATE {$\overline{\bf p} \Leftarrow \overline{\bf z}+\overline{\beta}\overline{\bf p}$}\;
\STATE {$\overline{\bf q} \Leftarrow \overline{\bf K}({\rm or} \overline{\bf A})\overline{\bf p}$}\;
\STATE {$\overline{\gamma} \Leftarrow (\overline{\bf p}, \overline{\bf q})$}\;
\STATE {$\overline{\alpha} \Leftarrow \overline{\rho}_{a}/\overline{\gamma}$}\;
\STATE {$\overline{\rho}_{b} \Leftarrow \overline{\rho}_{a}$}\;
\STATE {$\overline{\bf e} \Leftarrow \overline{\bf e}-\overline{\alpha}\,\overline{\bf q}$}\;
\STATE {$\overline{\bf u} \Leftarrow \overline{\bf u} + \overline{\alpha}\,\overline{\bf p}$}\;
\STATE {$i \Leftarrow i+1$}\;
\ENDWHILE
\end{algorithmic}
\end{minipage}
\end{algorithm}

% これは有用なHPC手法を組み合わせて開発されたものである．以下に各種の説明を加える．
%The solver is developed by combining several methods:
\begin{description}
%\item[{\it Adaptive Conjugate Gradient method}~\cite{ipcg}] %\mbox{}\\
\item[Adaptive Conjugate Gradient method~\cite{ipcg}:] 
% 本論文では代表的な反復法の一つである共役勾配法を用いる．
% 前処理の行列方程式${\bf M}{\bf z}={\bf r}$において，${\bf M}$が${\bf K}$に近ければ収束性が改善されることがわかっている．
In a standard preconditioner in the conjugate gradient method, a fixed matrix ${\bf M}^{-1}$ that is close to the inverse of ${\bf K}$ is used to improve the convergence of the iterative solver (i.e., ${\bf r} = {\bf M}^{-1}{\bf z}$).
% しかし、行列${\bf M}$を計算に用いると、使用メモリ量が増加してしまう．
% However, computation of ${\bf M}^{-1}$ is expensive and its memory footprint is large.
In the adaptive conjugate gradient method, the equation ${\bf K}{\bf z}={\bf r}$ is roughly solved instead of using a fixed matrix (${\bf M}^{-1}$), which in turn opens up room for improvement of the solver.
% そこで代替案として，${\bf K}{\bf z}={\bf r}$を解くことにより，${\bf z}$をある程度の誤差を許容して求める．
% アルゴリズム~\ref{algo}(a)の8行目から17行目が可変的前処理に該当する．
Here, lines 8--17 of Algorithm~\ref{algo}(a) correspond to the adaptive preconditioner, and a conjugate gradient method with 3 $\times$ 3 block Jacobi preconditioner is used for the inner loop solvers (Algorithm~\ref{algo}(b)).
% 以降、${\bf K}{\bf z}={\bf r}$を解くための反復をinner loop、元々の反復をouter loop と呼ぶ．
From here on, we refer to the iterations for solving the preconditioning equation as the inner loop, and the iterations of the original solver as the outer loop.
% inner loopは，反復回数と，残差ベクトルのノルムの双方によって終了判定がなされる．
The inner loops are terminated based on the maximum number of iterations and error tolerance.
%見かけ上はOuter loopの反復のたびにinner loopの反復回数が変わり，前処理の効果が変化しているように見なせる．
%このため，この前処理は可変的前処理と呼ばれる。
%本論文では内部反復、外部反復ともに共役勾配法が用いられる．
%\item[{\it Mixed Precision Arithmetic}]\mbox{}\\
\item[Mixed Precision Arithmetic:] 
% 本ソルバーでは，最終的な結果を倍精度にて得るため，Outer loopの計算は倍精度で行う．
Although double-precision variables are required for accurate calculation of the outer loop, 
% 一方で，Inner loopは外部反復の反復回数を抑えるために計算されるものであるため，高い精度での演算は必要ない．
the inner loops are required only to be solved roughly.
% そこで，単精度での演算が可能となる．
Thus, we use single-precision variables in the inner loops (denoted with bars in Algorithm~\ref{algo}).
By setting suitable thresholds in the inner solvers, we can shift computation cost from the outer loop to the inner loops, enabling double-precision results computed mostly with single-precision computation.
% これによってmemory footprintやデータ転送量，通信量が半減し，また見かけ上はキャッシュサイズを2倍として計算を行うことが可能となる．
This halves the memory footprint, memory transfer size, and communication size, and doubles the apparent cache size.
% アルゴリズム~\ref{algo}では，単精度の変数にバーが付与されている．
% \item[{\it Geometric/Algebraic Multi-grid method}]\mbox{}\\
\item[Geometric/Algebraic Multi-grid method:] 
% 内部反復自体の収束性の向上を試みるために，Multi-grid法を導入する~\cite{multigrid}．
We use a multi-grid~\cite{multigrid} for improving the convergence of the inner loops.
% オリジナルのモデルは，四面体二次要素モデルによるものであるため，Geometric Multi-grid法を用いることができる．
As the target problem is discretized with second-order tetrahedral elements, we first use a geometric multi-grid to coarsen the problem.
% すなわち，同じ領域に対して分解能の低い四面体一次要素モデルを同時に生成し，内部反復部の初期解の推定を行う．
Here, we use the same mesh but without edge nodes to construct the first-order tetrahedral element coarse grid, and we use the solution on this coarsened grid as the initial solution for the second-order inner loop.
% 前処理方程式${\bf z}={\bf K}^{-1}{\bf r}$は一次要素によるモデル上の${\bf z}_{1}={\bf K}_{1}^{-1}{\bf r}_{1}$に再び変換され，少ない反復回数によって計算が行われる．
%That is, the preconditioning equation ${\bf z}={\bf K}^{-1}{\bf r}$ is converted to ${\bf z}_{1}={\bf K}_{1}^{-1}{\bf r}_{1}$ of the first-order mesh that can be computed with less number of iterations.
% 以下では，二次要素によるモデルを用いた反復をinner loop level 0，一次要素によるモデルを用いた反復をinner loop level 1と呼ぶ．
From here on, we refer to the model with second-order elements as inner loop level 0, and the model with first-order elements as inner loop level 1.
% 一次要素によるモデルは二次要素によるモデルと比較して自由度が少ないため、必要な反復回数が多くなっても全体としての計算時間が軽減される．
As the degrees of freedom of the first-order model is smaller than that of the second-order model, we can expect speedup.
% 一方で，地殻変動など静的な問題に対しては，inner loop level 1の求解時間が支配的となる．
In the case of static crust-deformation problems, we can expect further speedup from further coarsening of the grids.
% そのため，Algebraic Multi-grid法を併せて用いる．
Here, we coarsen the first-order tetrahedral grid using the algebraic multi-grid method such that
% 一次要素モデルより荒いモデルを定義し共役勾配法を適用することによって，低周波成分の減衰を速め，inner loop level 1部の近似解を改善する．
% Here we apply the conjugate gradient solver to a model generated by coarsening the  model, such that 
low-frequency components of the solution can be resolved quickly using a conjugate gradient solver.
% このモデルはさらに自由度が低くなるため，全体の計算量の削減に大きく寄与する．
The degrees of freedom of this grid becomes further smaller, leading to further reduction in computing cost.
% 以下ではこのループをinner loop level 2と呼ぶ．
From here on, we refer to this as inner loop level 2.
% \item[{\it Element-by-Element method}]~\cite{EBE}\mbox{}\\
\item[Element-by-Element method~\cite{EBE}:]
% 連立一次方程式ソルバーの中で最も多くの計算コストを占める演算は疎行列ベクトル積である．
The most costly part of the solver consists of sparse matrix-vector products that are called in each iteration of the inner and outer conjugate gradient solvers.
% 本研究ではElement-by-Element (EBE) 法を利用した疎行列ベクトル積計算を適用する．
Here we use the element-by-element (EBE) method for computing sparse matrix-vector products.
% EBE法では式~\ref{EBEeq}のように，各要素剛性マトリクスとベクトルの積を加算することによって，行列ベクトル積の結果を得る．
In the EBE method, matrix-vector products are calculated by summing element-wise matrix-vector products as
\begin{equation}
\label{EBEeq}
{\bf f}=\sum_{i} {\bf Q}_{i}{\bf K}_{i}{\bf Q}_{i}^{T}{\bf u}.
\end{equation}
% ここで，{\bf K}は要素剛性マトリクス，{\bf Q}は全体ベクトルへのマッピング行列を表す．
Here, {\bf K}$_i$ indicate the element stiffness matrix and {\bf Q}$_i$ indicates the mapping matrix between local and global node numbers.
% 要素剛性マトリクスはメモリに保持せず，要素コネクティビティと座標，物性値の情報からその場で計算して作成する．
Instead of storing the element stiffness matrix in memory, it is computed every time a matrix-vector product is computed using nodal coordinates and material properties.
% これらの配列は繰り返し読み込まれるため，キャッシュを利用して計算することが可能で，疎行列ベクトル積をメモリバンド幅律速から演算律速の計算へと移行することが可能となる．
As {\bf u} and coordinate information are read many times during the computation of Eq. (\ref{EBEeq}), it can be stored on cache. This enables shifting the memory bandwidth load to an arithmetic load in sparse matrix-vector multiplication.
% これは昨今の演算性能に対してメモリバンド幅が少ないアーキテクチャに対して特に高い効果を発揮する．
This is especially effective when targeting recent architectures with high arithmetic capability per memory bandwidth capability. 
% また疎行列格納による行列ベクトル積と比較すると，全体のメモリアクセス量も軽減される上に，コアごとに同数の要素を割り当てることによって容易にロードバランスを保つことが可能である．
% When compared with standard sparse matrix-vector product kernels reading global matrices from memory,
In addition to the reduction in memory transfer, we can also expect improvement in load balance by allocating the same number of elements per core.
% 式~\ref{EBEeq}の計算においては，SIMDによる並列計算を活用するために，マルチコアカラーリングとSIMDバッファリング手法を用いている．
In the CPU-based implementation, multi-coloring and SIMD buffering is used to attain high performance on multi-core SIMD-based CPUs.
% このEBE計算は，Outer loopと，inner loop level 0, 1 の疎行列ベクトル積に適用される．
This EBE computation is applied to the outer loop, inner loop level 0, and inner loop level 1.
% inner loop level 2については，通常の疎行列ベクトル積と比較して計算コストが大きくなるため，EBEによる計算を行わない．
As the matrix for inner loop level 2 is algebraically generated and thus EBE method cannot be applied, we read the global matrix from memory stored in 3 $\times$ 3 block compressed row storage format.
% 疎行列をメモリ上に保存して読み込む必要性があるが，自由度はオリジナルのものと比較すると極めて小さいため，footprintへの影響は小さいものと考えられる．
As the model for inner loop 2 is significantly smaller than the original second-order tetrahedral model, the memory footprint for storing level 2 models is expected to be small.
\end{description}

% 以上の手法は単精度演算やマルチグリッド法をはじめとして，計算量・データ転送量が軽減されるように設計されている．
In summary, the method above is designed to reduce computation cost and data transfer size with good load balancing through the combination of several methods.
% このため，GPUに対しても適したアルゴリズム構成と考えられる．
Such properties are also expected to be beneficial for GPUs as well.
% 次章ではこのソルバーに対して，実際にOpenACCによってGPU計算を導入し，その性能を確認する．
In the next section, we explain porting of this solver using OpenACC and measure its performance.

\section{Proposed Solver for GPUs using OpenACC}
\label{sct3}
Compared with CPUs, GPUs have relatively smaller cache sizes and tend to be latency bound for computation with random data access.
Therefore, algorithm and implementation to attain optimal performance in GPU differ from the base algorithm for CPU-based computers.
We modify the solver algorithm such that random memory access is reduced, and we port this modified algorithm solver to GPUs using OpenACC.
Subsequently, we first explain the algorithm modification and then the details of porting with OpenACC.

\subsection{Modification of Algorithm for GPUs}
% OpenACCを導入するにあたって，GPUの性能を十分に活かせるよう，適したアルゴリズムへの拡張を行う．
We first update the solver algorithm to suit the GPU architecture.
% 想定している問題は同一の剛性マトリクスに対して多数のベクトルを入力し，多数回の連立一次方程式の求解を行う．
The target application requires solving many systems of equations with the same stiffness matrix but different right-hand side input vectors.
% そこで本研究では，複数回の計算をまとめることによって，計算性能を向上させることを考える．
Thereby, we improve performance by conducting multiple computations simultaneously.
% 解析内の計算時間の大部分を占めているのが疎行列ベクトル積である．
% Sparse matrix-vector products accounts for large proportion of the whole computation cost in the analysis.
% EBE計算は乗算されるベクトルのロードストアが律速となっている．
Since the performance of the most costly EBE kernel is bound by loading and storing of data,
% このベクトルへのメモリアクセスの不規則性を軽減することができれば，単純にポートするよりも高い性能を得ることができると期待される．
we can expect significant performance improvement by reducing the irregularity of memory accesses.
% そこで,複数のベクトルを用意し，それぞれに対して同一のマトリクスを一度にかけ合わせるような計算方法をとる．
Based on this idea, we solve multiple systems of equations simultaneously by multiplying the same element stiffness matrix to multiple vectors at the same time.
% これによって，ベクトルの本数分だけ連続にメモリアクセスを行うことが可能となるため，単純に行列×ベクトル1本を繰り返すよりも高速に計算が可能となる．
This approach enables coalesced memory access for the number of vectors, leading to a shorter time to solution than repeating multiplication of a matrix and a single vector multiple times.
% 本研究では，16本のベクトルを用意し，並列に計算を進める．
In this paper, we solve 16 systems of equations in parallel
% すなわち，我々は${\bf K}\left[ {\bf u}_1, {\bf u}_2, ..., {\bf u}_{16} \right]^{T}=\left[ {\bf f}_1, {\bf f}_2, ..., {\bf f}_{16} \right]^{T}$として連立一次方程式の求解を行う．
% That is, we solve systems of linear equations as 
(${\bf K}\left[ {\bf u}_1, {\bf u}_2, ..., {\bf u}_{16} \right]^{T}=\left[ {\bf f}_1, {\bf f}_2, ..., {\bf f}_{16} \right]^{T}$).
% 各種のループの収束については，各残差ベクトルのノルムの最大値を用いることによって判定を行う．\\
% このアルゴリズムの変更によって，各計算は2重のループを含んだ構造となり，最も内側のループがベクトル本数のループとなる．
This modification changes all of the computational loops into nested loops, with the inner loop having a loop length of 16.
The maximum values for the errors in the 16 residual vectors are used for judging the convergence of each loop.

\subsection{Introduction of OpenACC}
%修正を加えたアルゴリズムに対してOpenACCの導入を行う．
We introduce OpenACC to the modified algorithm.
% ソルバー部分全体をGPUのポーティング対象として，計算時間を短縮させる．\\
Here, the solver part is ported to GPUs to reduce the application runtime.
% OpenACCを導入するにあたっては，データ転送を最小化したうえで，各計算をGPUへ移植する必要がある．
For high performance, we first need to maintain data transfer at a minimum and then conduct all computation on the GPUs.
% そこでこの順にOpenACCの実装方法について説明する．

\subsubsection{Control of Data Transfer}
% OpenACCは明示的に指示しなければ並列計算領域前後で使用するデータすべてに対して自動で転送を行うため，計算性能が著しく低下する．そこでデータの保持・転送を管理する指示行の挿入が必須である．
Unless explicitly specified otherwise, OpenACC automatically transfers all data necessary for GPU computation between the host memory and the GPU device memory every time a kernel is called. This data transfer seriously degrades performance; thus, we insert directives to control the data transfer.
% データ転送が必要となるのは，MPIによる袖通信と，各ループにおける収束判定のみである．
In the solver, data transfer is necessary only for MPI communication and checking the convergence of each loop.
% これら以外のデータ転送を防ぐため，data指示行のpresentおよびcopy, copyin, copyoutオプションによってその都度転送の要不要を明示する．
We use the {\tt present} option in the {\tt data} directive of OpenACC for other parts of the solver to eliminate unnecessary data transfer.
% MPIを用いた袖領域の通信についてはGPU directを用いた高速な転送を導入する．
For the MPI communication part, we use GPU Direct, which enables MPI communication without routing through the host memory system.
This is enabled by 
% OpenACCについては，デバイス側のメモリを使用することを宣言する指示行が用意されているため，これらの指示行をMPI通信の前後に挿入することによって，GPUのデバイスメモリからCPUのメモリを経由せずに直接データの送受信が可能となる．
inserting OpenACC directives before and after MPI communication that declares the use of device memory.

\subsubsection{Porting of Each Kernel}
\begin{figure}[tb]
%\begin{minipage}{0.51\hsize}
\begin{center}
 \includegraphics[clip,scale=0.53]{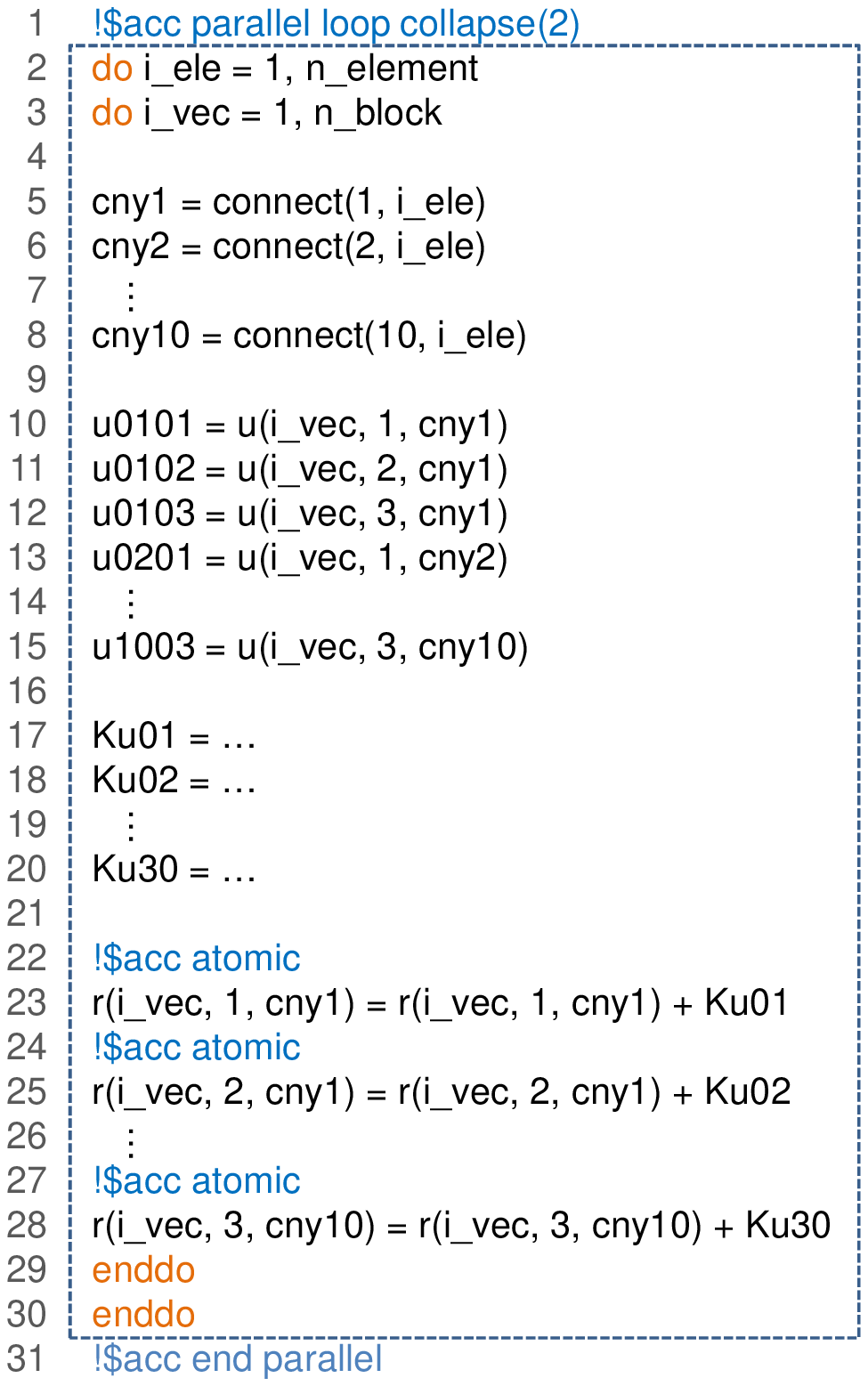}
      \end{center}
     \caption{EBE kernel for multiple vectors on GPUs.}
     \label{porting2}
%\end{minipage}
\end{figure}

% 続いて，各カーネルに並列化を指示するためのloop指示行と，それぞれの計算に対応するオプションを加えることによるGPUへのポートを行う．
Next we port each kernel by using the {\tt loop} directives with suitable options.
% 図~\ref{porting2}はEBE計算におけるFortranでのポーティング例を示している．
Figure~\ref{porting2} shows a porting example of the EBE kernel with multiple vectors in Fortran.
%EBE計算をはじめとして本ソルバー内のカーネルはいずれも，本来のループの内側にベクトルの本数分のループを合わせた2重のループを持つ．
Each of the kernels in the solver has a nested loop with an inner loop length of 16.
% 最も内側のループ長は長くないため，collapse指示行を加えることによって，2重ループをまとめて1つとして容易に並列化することが可能となる．
The length of the inner loop is not large; thus, we collapse these nested loops by adding {\tt collapse} options in the {\tt loop} directives.
% OpenACCの仕様によりcollapse指示行は対象となるループが隣接していないと使用することができないため，EBE計算の例では要素コネクティビティの配列は重複して読み込まれる．
In the current specification of OpenACC, target loops must be adjacent to each other for collapsing loops. 
Thus, parts of the kernel must be computed redundantly (e.g., the node connectivity array is read redundantly in Fig.~\ref{porting2});  however, collapsing of the loops enables
% 複数ベクトル生成時には1次元目にベクトルにあたる変数を割り当て，一部メモリの連続アクセスを可能とする．
% The first argument is corresponding to the number of vectors and this partially enables 
coalesced memory accesses for the vectors leading to higher computing performance. 
% GPUではSIMT計算を自動的に行ってくれるため，CPUのように明示的にSIMT計算が適用されるようにコードを書く必要はない．
In GPU computation, SIMT computation is applied automatically; thus we do not have to designate parallel code explicitly as in SIMD computation in CPUs.
% 全体ベクトルへの足しこみに対してはatomic演算に対応する指示行を挿入する．
We insert  {\tt atomic} directives for adding thread-wise temporal variables to the resulting vector.
% 既往の研究により，カラーリングを用いて計算の順番を制御するよりもデータの局所性を高く保って計算を行うことができることがわかっている．~\cite{waccpd2016}
Our previous study has shown that atomic operations attain higher performance than reordering the elements to avoid a data race using the coloring method \cite{waccpd2016}, because atomic operations can retain data locality of nodal data (i.e., {\bf u} and nodal coordinate information) and thus utilize the L2 cache more efficiently.
% その他のベクトルの定数倍，加算，減算などについても同様に，loop指示行およびcollapseオプションを挿入するだけで計算がGPUによって高速に処理される．

\begin{figure}[tb]
%\begin{minipage}{0.45\hsize}
\begin{center}
 \includegraphics[clip,scale=0.53]{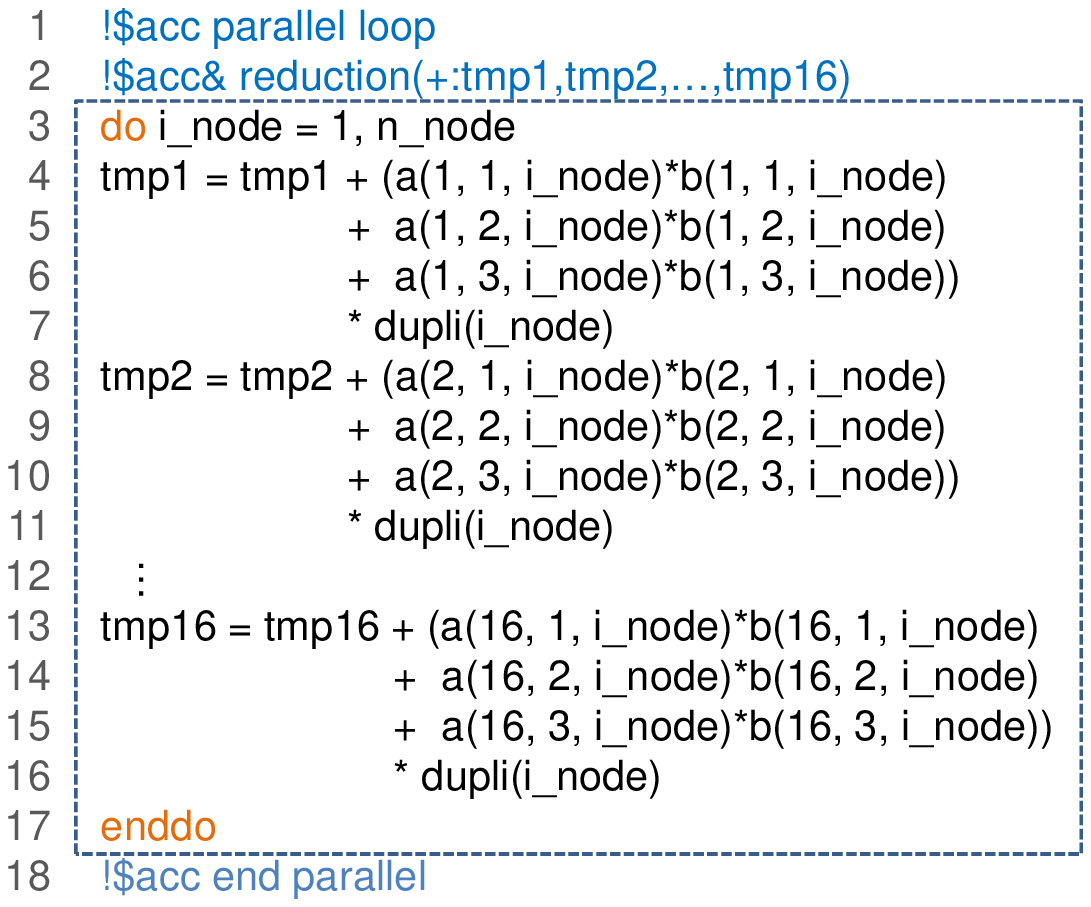}
 \end{center}
     \caption{Vector inner product kernel for multiple vectors on GPUs.}
     \label{porting3}
%\end{minipage}
\end{figure}

Although other calculations, such as multiplication, addition, and subtraction of vectors, can also be computed on GPUs by adding {\tt loop} directives, we must take care when porting the inner product kernel with multiple vectors.
% 内積計算では本来loop指示行にreductionオプションを加えるのみで容易に計算が可能である．
In the case of inner vector products with a single vector, we can directly port the CPU code by inserting the  {\tt reduction} option in  {\tt loop} directives.
% ただし複数ベクトル時の計算については注意が必要となる．
However, 
% 現状ではOpenACCの制約上，reductionが作用するのはスカラーのみで，配列を対象とした縮約が指示できない．
the reduction option in OpenACC is available for scalars but not for arrays.
% これは，OpenACCの現在の仕様では内側のループを並列化することは不可能であることを意味している．
Thus, the innermost loop cannot be parallelized directly in the present specification of OpenACC.
% そこで計算に当たってはスカラーをベクトル本数分だけ用意し，節点数だけループを回す．
Thereby, we allocate scalars corresponding to each of the multiple vectors and compute the reduction of these scalars in a single loop
% ポーティング例は図~\ref{porting3}のようにあらわされる．
(Fig.~\ref{porting3}).
% GPUの各スレッドがそれぞれ離れた位置を参照することになるため，性能が低下することになることが考えられる．
%Each thread in a GPU refers to separate components of arrays; thus the performance of inner vector product is expected to be decreased.
In this case, memory access becomes strided, possibly leading to performance decrease when compared with the single vector type inner product kernel.
Adding collapse options with reduction options for arrays in OpenACC that enable contiguous memory access might be beneficial in this case.
% 一方で，内積の計算時間は疎行列ベクトル積の計算時間と比較すると比較的小さく，計算時間の影響も疎行列ベクトル積と比較すると小さいものと予想される．
% On the other hand, the computation time of inner vector product is thought to be much smaller than that of sparse matrix-vector product and this limitation in parallelization has small impact on the whole computation time.

% 以上でGPU計算が可能となるが，性能を十分に得るためには，OpenACC内での並列性を決定するパラメータについての検討が必要となる．
Examining the parameters in OpenACC defining parallelism of computation is important for exhibiting high performance.
% OpenACCではgang, worker, vectorの3階層によって並列化の粒度が決定される．
In OpenACC, three hierarchies of  {\tt gang}, {\tt worker} and {\tt vector} determine the granularity of parallelization.
% NVIDIA GPUとの計算上の構造の関連付けとしては，gangがthread block, vectorがthreadに対応している。
The parameter {\tt gang} corresponds to a thread block in an NVIDIA GPU, and {\tt vector} corresponds to a thread. 
%Element-by-Element計算を例に挙げると,GPUの1スレッドが1要素の計算を担当することで最も高い性能を得ることが可能となる．
For instance, in EBE kernels, we must assign one thread per element to attain their optimal performances on GPUs.
%明示的に指示を加えない場合，要素内の計算に含まれる小さなループに複数のスレッドが割り当てられるなど，意図しない並列化が実行されるため，%これらのパラメーターを適切な位置に挿入することが不可欠となる．
Without any instructions, threads can be unintendedly mapped; thus these two options, gang and vector, must be inserted in appropriate places explicitly.
% OpenACCでは，明示的に設定しない場合，vectorの長さ(NVIDIA GPUでのブロックサイズに相当する)はコンパイラが自動的に決定する．
The length of {\tt vector}, which corresponds to the block size in NVIDIA GPUs, is automatically determined by the compiler.
% 多くの場合性能に大きな影響が出ることは少ないが，特に全体の計算性能を大きく左右するカーネルについてはそれぞれのカーネルについて適切なパラメーターを試行・確認する必要がある．
In most cases, these parameters have little impact on performance; however, we searched for optimal parameters for core kernels in the solver to attain optimal performance.
% 多くのレジスタを用いるEBEカーネルについては，ベクトル長を32と設定することができ，これによって異なるwarpとの同期が不必要となるため，もっともよいパラメータとして計算が可能になると考えられる．
For EBE kernels, in which the usage of registers rather than the block size has the largest impact on the number of working threads, we can set their lengths of vector to 32. This saves the need for synchronization in the block among different warps; thus this parameter is expected to be the most optimal one.
%このように開発されたソルバーについて，以下の章では性能測定を行う．

\section{Performance Measurements}
\label{sct4}
% 本章では提案するOpenACC版ソルバーについての性能計測を行い，提案手法の有効性を示す．
In this section, we show the effectiveness of the developed solver through performance measurements.

\begin{figure}[tb]
\begin{center}
 \includegraphics[clip,scale=0.55]{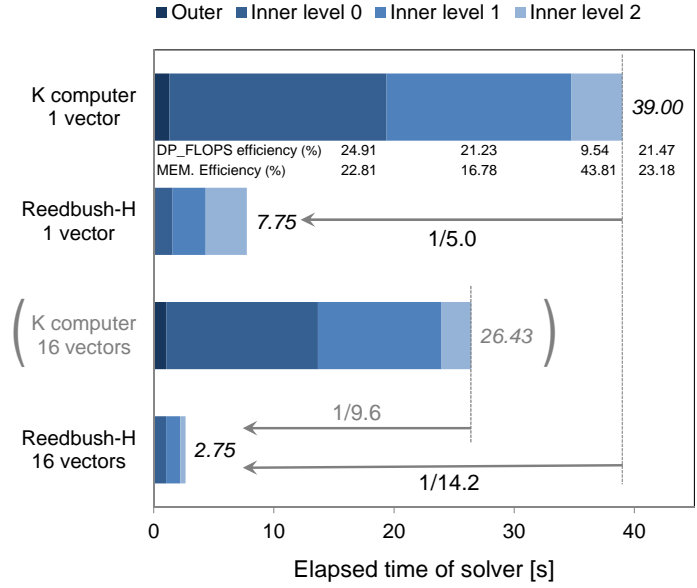}
     \caption{Performance comparison of the entire solver. Here, the computation time when using 16 vectors is divided by 16 and converted per vector.}
     \label{measure12}
\end{center}
\end{figure}

% 初めに，京コンピューターとNVIDIA P100 GPUとを用いた性能計測を行う．
We first compare the performance of the ported solver on NVIDIA P100 GPUs with that of the base solver on the K computer.
% P100 GPUを用いた計算機環境としては，東京大学情報基盤センターが運用しているReedbush-Hを用いる．
Here we use Reedbush-H of the Information Technology Center, The University of Tokyo for a computation environment with P100 GPUs~\cite{reedbush}.
The K computer is a massively parallel CPU-based supercomputer system at the Advanced Institute of Computational Science, RIKEN.
% それぞれの計算機環境は表~\ref{env_KandP}にまとめられている．
The computation environments are summarized in Table~\ref{env_KandP}.
% なお，本論文では，inner loopの収束判定条件は表~\ref{loopconfig}のように設定して計算を行う．In this paper, the maximum iteration and tolerance thresholds for inner loops are configured as in Table~\ref{loopconfig}.
% これらの閾値はinner loopの計算時間およびouter loopの収束性に大きく影響を与えるものと考えられる．
% Outer loopは誤差が10^{-8}を下回るまで計算を続ける．
We continue to compute until the residual error in outer loop goes below $10^{-8}$. The maximum iteration and tolerance thresholds for inner loops greatly affect the convergence of the outer loop and whole computation time.
%　問題によっても収束性が異なるため適したパラメーターを一意に決定することは難しいが，今回は性能測定にあたって経験的にソルバーの計算時間の全体が短縮されるようにそれぞれのパラメーターを決定している．
In this paper, these parameters are empirically configured as in Table~\ref{loopconfig} so that the computation time for the whole solver is reduced.
% また使用するモデルは自由度が125,177,217，要素数が30,720,000となる有限要素モデルを用いる．
For performance measurement, we use a finite-element model with 125,177,217 degrees of freedom and 30,720,000 second-order tetrahedral elements.
% モデルは2層からなっており，各層の物性値は表~\ref{materialconfig}のように設定している．
The material properties of the two-layered model are shown in Table~\ref{materialconfig}.
% CGループの計算時間について，京コンピューターとの比較を行ったところ，それぞれの計算時間は図~\ref{measure12}のように表された．
The computation time in the conjugate gradient loop is shown in Fig.~\ref{measure12}.
% この図に関してはベクトル1本あたりの計算時間によって比較を行っている．
From the figure, we can see that the base solver attains 21.5\% of peak FLOPS on the K computer system, which is very high performance for a low-order finite-element solver.
Compared to this highly tuned CPU solver implementation,
% This figure also compares computation time per vector.
% 単純に移植を行っただけで5.0倍の高速化が実現しているが，この数値は
we confirmed that direct porting of the original solver without algorithm changes enabled 5.0 times speedup.
% from the K computer.
% ピーク性能の差を考慮するとP100 GPUの性能を十分に活用できていないと評価できる．
% When considering the difference between the peak FLOPS and memory bandwidth capabilities of the two systems, we can expect room for improvement.
% 一方で16本のベクトルをまとめて高効率な計算を導入することによって，ベクトル1本あたりの計算時間は1/2.82に短縮された．
By using the proposed method solving 16 vectors simultaneously in GPUs, the solver was accelerated further by 2.82 times per vector from the directly ported solver.
% 京コンピューターと比較すると，これは14.2倍の高速化となっている．\\
This leads to 14.2 times speedup with respect to the base solver on the K computer.
%　比較のため，京コンピュータでも16本ベクトルを用いてソルバーの計算を行ったところ，図に示されるように，その計算時間はベクトル1本あたり26.43秒で，オリジナルのアルゴリズムのものと比較するとその高速化効率は1.48倍であった．GPU計算機上での複数ベクトル導入による高速化倍率は2.82倍であることから，ランダムアクセス量を軽減させるアプローチはCPUにも同様に有効であるが，特にGPU計算で効果を発揮することが確認できる．\\
For comparison, we measured the computation time using 16 vectors on the K computer. As shown in Fig.~\ref{measure12}, we attained 1.48 times speedup with regard to the original solver using single vector. Considering the speedup ratio is 2.82 between using 16 vectors and single vector on Reedbush-H, we can confirm that we have attained higher performance in P100 GPUs by introduction of dense computation, which is more effective for GPU computation.
%16本ベクトルを共に用いた場合のKコンピューターとReedbush-Hの性能差は9.6倍となっている．
The speedup ratio in using 16 vectors is 9.6 between Reedbush-H and K computer.
% As seen in the Table~\ref{env_KandP}, there is 11.4 times and 41.4 times difference in the peak memory bandwidth and peak DP FLOPS between the systems.
In standard finite-element solvers using sparse matrix storage formats, the expected speedup will be near the peak memory bandwidth ratio, which is 11.4 times in this case.
However, for practical cases, the computation includes random accesses that severely degrade GPU performance; thus the speedup ratio is assumed to get much less than the peak bandwidth ratio.
%the problem size will be too small for the 16 GB GPU memory to attain such high performance for standard solvers.，
Thus, we can see that the 9.6 times speedup attained is reasonable performance.
%これはElement-by-Element法を疎行列ベクトル積演算に用いているため，メモリバンド幅律速の計算をキャッシュのバンド幅律速へと移行することができていることが大きな一因としてあげられる．
This speedup ratio is mainly due to the introduction of EBE multiplication, which has changed global memory bandwidth bound computation into cache memory bandwidth bound.
% また，複数ベクトルを用いたことによる各カーネルの性能の変化についても確認を行った．
We examine the cause of the performance improvement by checking the speedup of each kernel in 
% 表~\ref{kernelperformance}はGPUを用いた際のソルバー部分の計算時間について，各種カーネルの計算時間を取り出したもので，それぞれベクトル1本あたりの計算時間に換算されている．
Table~\ref{kernelperformance}.
% AMG部分の疎行列格納形式を用いた疎行列ベクトル積については，巨大な剛性マトリクスの読み込みが支配的となっている．
% そのためベクトルの本数が増えても大きく計算時間が上昇することがなく，計算効率が大きく向上している．
As the sparse matrix-vector product in inner loop level 2 is bound by reading the global matrix from memory, the total computation time is nearly constant regardless of the number of vectors multiplied.
Thus, the efficiency of computation is significantly improved when 16 vectors are multiplied simultaneously.
% また，EBE計算ではメモリアクセスのランダム性が大きく抑えられるため，いずれのカーネルも性能が1.6倍から1.7倍向上していることが確認できた．
The reduction in random memory access in EBE kernels leads to performance improvement by 1.6--1.7 times.
% 内積計算についてはメモリアクセスにストライドが発生するため，性能が落ちている．
Although the performance of the inner vector product kernel decreased as a result of strided memory access,
% 一方で，疎行列格納形式による疎行列ベクトル積およびEBE計算の高速化の影響が大きく，ソルバー全体でみると計算効率は大きく上がっていることが確認できた．
the acceleration of sparse matrix-vector products and EBE computation has a profound effect leading to performance improvement of the entire solver.

\begin{table}[tb]
\caption{Comparison of hardware capabilities of K computer and Reedbush-H.}
\label{env_KandP}
\begin{center}
\begin{tabular}{lrr}
 & K computer & Reedbush-H (Tesla P100)\\
\hline 
\# of nodes & 20 & 10 \\
\multirow{2}{*}{CPU/node} & 1 $\times$ eight-core  & 2 $\times$ eighteen-core \\
& SPARC64 VIIIfx & Intel Xeon E5-2695 v4  \\
Accelerator/node & - & 2 $\times$ NVIDIA P100 \\
\# of MPI processes/node & 1 & 2 \\
hardware peak DP & \multirow{2}{*}{128 GFLOPS} & \multirow{2}{*}{5.30 TFLOPS (GPU only)} \\
FLOPS /process &&\\
Bandwidth/process & 64 GB/s & 732 GB/s (GPU only) \\
\multirow{3}{*}{Interconnect} & Tofu (4 lanes $\times$  & PCIe Gen3 $\times$ 16  \\
 & 5GB/s in both directions) &  + NVLink (20 GB/s) $\times$ 2\\
 & & + InfiniBand FDR 4 $\times$ 2 \\
\multirow{2}{*}{Compiler} & Fujitsu Fortran & \multirow{2}{*}{PGI compiler 17.5} \\
&Driver Version 1.2.0& \\
\multirow{2}{*}{Compiler option} & -Kfast,openmp,parallel,ocl &-ta=tesla:cc60,loadcache:L1\\
&& -acc -Mipa=fast -fastsse -O3 \\ 
 MPI & custom MPI & OpenMPI 1.10.7 \\
\hline 
\end{tabular}
\end{center}
\end{table}

\begin{table}[tb]
\caption{Error tolerance $\overline{\epsilon}_{0-2}$ and maximum iteration $N_{0-2}$ used for our solver for solving measurement models and application problems.}
\label{loopconfig}
\begin{center}
\begin{tabular}{lrr}
Inner loop \ \ & \ \ Error tolerance& \ \ Maximum iteration\\
\hline
level 0 & 0.1 & 30\\
level 1 & 0.05 & 300\\
level 2 & 0.025 & 3,000\\
\end{tabular}
\end{center}
\end{table}

\begin{table}[tb]
\caption{Material properties of performance measurement models. $V_p$, $V_s$, and $\rho$ indicate primary wave velocity, secondary wave velocity, and density, respectively.}
\label{materialconfig}
\begin{center}
\begin{tabular}{lrrr}
Layer \ \ \ & $V_p$ (m/s)& $V_s$ (m/s)& $\rho$ (kg/m$^3$)\\
\hline
1 & 1,600 & 400 & 1,850\\
2 & \ \ \ 5,800 & \ \ \ 3,000 & \ \ \ 2,700\\
\end{tabular}
\end{center}
\end{table}

\begin{figure}[tb]
\begin{center}
 \includegraphics[clip,scale=0.55]{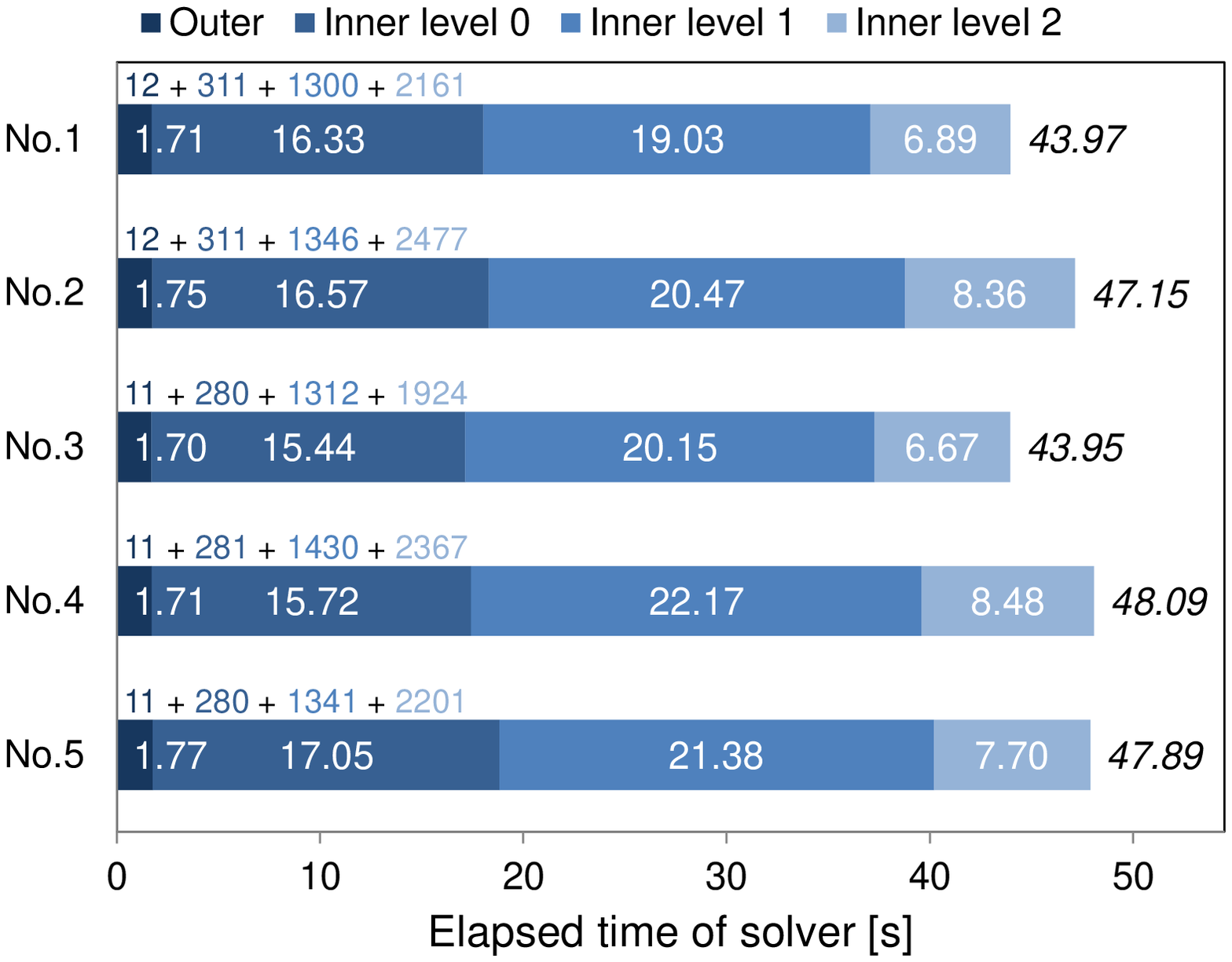}
     \caption{Performance in weak scaling. The numbers of iterations for the outer loop, inner loop level 0, inner loop level 1, and inner loop level 2 are written in the insets.}
     \label{scaling}
\end{center}
\end{figure}

\begin{table}[tb]
\caption{Performance of main kernels in Reedbush-H.}
\label{kernelperformance}
\begin{center}
\begin{tabular}{l|rr|r}
 \multirow{3}{*}{Kernel} & \multicolumn{2}{c|}{Elapsed time per vector (s)} & \ \ \ \multirow{3}{*}{Speedup}\\ 
%   & \multicolumn{2}{c|}{per vector (s)} & \\
  & \ \ 1 vector  &  \ \ \  16 vectors & \\ \hline
SpMV ($\overline{\bf A}_{2} \overline{\bf u}_{2}$) & 1.465&  0.091 & 16.10 \\
2nd order EBE (${\bf K} {\bf u}$) &  0.044  &   0.025 & 1.78 \\
2nd order EBE ($\overline{\bf K} \overline{\bf u}$) &  0.687 &  0.401 & 1.71 \\ 
1st order EBE ($\overline{\bf K}_{1} \overline{\bf u}_{1}$) & 0.948 & 0.584 & 1.62 \\
Inner product ($\overline{\bf p} \cdot \overline{\bf q}$)& 0.213 & 0.522 & 0.41 \\ \hline \hline
Total time of the solver & 7.75 & 2.75 & 2.82
\end{tabular}
\end{center}
\end{table}
% 次に，弱スケーリングの測定を行うことによって，並列化性能の確認を行う．
We next check the parallelization efficiency by measuring weak scaling.
% Reedbush-Hの全系を用いて，ソルバーの計算時間を測定する．
Here, we measure the elapsed time of the solver using the full Reedbush-H system with 
% 全系では240枚のP100 GPUが使用可能である．
%full nodes of Reedbush-H have 
240 P100 GPUs. 
% 使用GPU枚数，自由度，四面体要素数表~\ref{modelconfig}にまとめる．
The number of GPUs, degrees of freedom, and the number of elements of the models are shown in Table~\ref{modelconfig}.
% ここで，モデルNo.1はK computerとの比較で用いた有限要素モデルと同一のものである．
Here, model No.1 is the same model as used in the performance comparison with the K computer.
% スケーリングの測定に当たっては，PCGEを用いた場合の反復回数の確認も併せて行った．PCGEはElement-by-Element法をベースとした標準的な共役勾配法で，アルゴリズム~\ref{algo}(a)から可変的前処理部分に該当する8行目～17行目を除いたものに相当する．
To assure that the convergence characteristics of the models are similar, we compared the number of iterations required for convergence of a standard conjugate gradient solver with a 3 $\times$ 3 block Jacobi preconditioner (from here on referred to as PCGE). PCGE corresponds to Algorithm~\ref{algo}(a) without the adaptive conjugate gradient preconditioner part (lines 8--17).
% 表~\ref{modelconfig}ではPCGEでの反復回数がほぼ一定となっており，弱スケーリングの測定に適したモデルセットであることが確認できる．
From Table~\ref{modelconfig}, we can see that the number of iterations in PCGE is nearly constant, and thus this model set is suitable for measuring weak scaling.
% AMG使用時の計算時間および各loopにおける反復回数を図~\ref{scaling}に示す．
Figure~\ref{scaling} shows the elapsed time of the developed solver and the total number of iterations required for convergence.
% AMG部分の計算回数にばらつきがあるため，計算時間は反復回数に従って増減している．
Although there are slight fluctuations in the number of iterations of the inner loops,
% 全体としてはおおよそスケールしており，並列化効率に関してもよい性能であると評価することができる．
the computation time is roughly constant up to the full system.

\begin{table}[tb]
\caption{Model configuration for weak scaling in Reedbush-H.}
\label{modelconfig}
\begin{center}
%\begin{tabular}{p{3em}|p{4.5em}p{6.5em}p{6.5em}p{6em}p{7em}}
\begin{tabular}{lrrrrr}
 Model & ~  \# of GPUs&  Degrees of freedom (DOF) &~DOF per GPU & ~\# of elements &~PCGE iterations  \\
\hline
No.1 & \hfill 20 & \hfill 125,177,217 & \hfill 6,258,861 & \hfill 30,720,000 & \hfill 4,928 \\
No.2 & \hfill 40 & \hfill 249,640,977 &\hfill  6,241,024 & \hfill 61,440,000 & \hfill 4,943 \\
No.3 & \hfill 80 & \hfill 496,736,817 & \hfill 6,209,210 & \hfill 122,880,000 & \hfill 4,901 \\
No.4 & \hfill 160 & \hfill 992,038,737 & \hfill 6,200,242 & \hfill 245,760,000 &\hfill 4,905 \\
No.5 & \hfill 240 & \hfill 1,484,953,857 & \hfill 6,187,308 & \hfill 368,640,000& \hfill 4,877\\
\end{tabular}
\end{center}
\end{table}

\begin{figure}[tb]
\begin{center}
 \includegraphics[clip,scale=0.55]{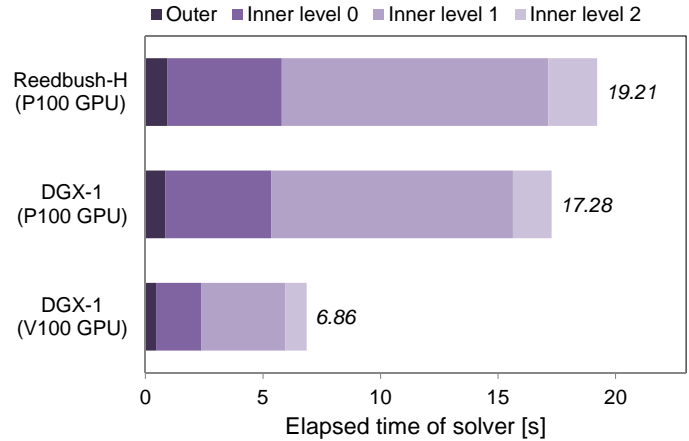}
     \caption{Performance comparison of the entire solver on Reedbush-H, DGX-1 (P100), and DGX-1 (V100).}
     \label{measure3}
\end{center}
\end{figure}

% ここまでP100 GPUを用いて手法の有効性を確認してきたが，最新のアーキテクチャを用いても十分な性能が得られることを確認する．
Finally, we check the effectiveness of the developed solver on the latest Volta GPU architecture.
Here, we compare performance of four Reedbush-H nodes with eight P100 GPUs, a DGX-1 with eight P100 GPUs, and a DGX-1 with eight V100 GPUs \cite{DGX1} (Table~\ref{env_PandV}).
The target model size is 38,617,017 degrees of freedom and 9,440,240 tetrahedral elements, almost filling the 16 GB device memory of the eight P100 and V100 GPUs on each system.
From Fig.~\ref{measure3}, we can see that the elapsed time has decreased from 19.2~s to 17.3~s when DGX-1 (P100) is used.
This performance difference may be attributed to the inter-node InfiniBand communication between the four Reedbush-H nodes in contrast to the intra-node communication inside a single DGX-1.
In the comparison of the P100 and V100 versions of DGX-1, the elapsed time has decreased from 17.3~s to 6.86~s.
This corresponds to 2.52 times speedup, higher than the 1.23 times increase in hardware peak memory bandwidth.
%この性能差は，キャッシュ部分におけるアーキテクチャの改善が大きな要因として考えられる．
Architectural improvements for caches contribute to this speedup ratio.
Volta GPU has 128 kB of combined L1 cache/shared memory per SM and 6 MB of L2 cache per GPU, which are 5.3 times and 1.5 times larger than L1 and L2 cache of Pascal GPU, respectively.
In the solver, random memory accesses in sparse matrix-vector multiplications is one of bottlenecks. Larger cache size in V100 GPU is thought to reduce memory bandwidth demand and improve performance of these kernels, including atomic addition part. Thereby it is inferred that these improvements in the hardware result in a speedup ratio more than the peak memory bandwidth ratio or the double-precision peak performance ratio.
%We measured device memory throughput for  Throughput for read/write are 44/17 GB/s on P100 GPUs in DGX-1 and 190/71 GB/s on V100 GPUs in DGX-1.}\textcolor{red}{To confirm the bandwidth of each cache, we measured L1/L2 throughput for Element-by-Element kernel in inner level loop 1, which is one of the most computationally expensive parts. Throughput of L1, L2 (read), and L2 (write) are 393 GB/s, 447 GB/s, and 314 GB/s on P100 GPUs in DGX-1 and 2,165 GB/s, 1,353 GB/s, and 748 GB/s on V100 GPUs, respectively. Each bandwidth has increased at least 2 times; thus we can conclude that this improvement results in a speedup ratio more than the peak memory bandwidth ratio or the double-precision peak performance ratio. }

\begin{table}[tb]
\caption{Comparison of hardware capabilities of Reedbush-H, P100 DGX-1, and V100 DGX-1. Latest compilers available in each environment are used.} 
\label{env_PandV}
\begin{small}
\begin{center}
\begin{tabular}{lrrr}
 & Reedbush-H (P100)& DGX-1 (P100) & DGX-1 (V100)\\
\hline 
\# of nodes & 4 & 1 & 1\\
\multirow{2}{*}{CPU/node} & 2 $\times$ eighteen-core & 2 $\times$ twenty-core & 2 $\times$ twenty-core \\
 & \ \ Intel Xeon E5-2695 v4  & \ \ Intel Xeon E5-2698 v4 & \ \ Intel Xeon E5-2698 v4  \\
Accelerators/node & 2 $\times$ NVIDIA P100 & 8 $\times$ NVIDIA P100 & 8 $\times$ NVIDIA V100 \\
MPI processes/node & 2 & 8 & 8 \\
GPU memory & \multirow{2}{*}{16 GB} & \multirow{2}{*}{16 GB} & \multirow{2}{*}{16 GB} \\
 size/process & \\
GPU peak DP & \multirow{2}{*}{5.3 TFLOPS} & \multirow{2}{*}{5.3 TFLOPS}& \multirow{2}{*}{7.5 TFLOPS} \\
FLOPS/process &&\\
GPU memory & 732 GB/s & 732 GB/s & 900 GB/s \\
bandwidth/process & \\
\multirow{3}{*}{Interconnect} &  InfiniBand FDR 4 $\times$ 2 & InfiniBand EDR $\times$ 4 & InfiniBand EDR $\times$ 4  \\
 & + PCIe Gen3 $\times$ 16  &  + NVLink &  + NVLink \\
 & + NVLink& & \\
Compiler & PGI compiler 17.5 & PGI compiler 17.9 & PGI compiler 17.9 \\
\multirow{4}{*}{Compiler option} & -ta=tesla:cc60 & -ta=tesla:cc60 & -ta=tesla:cc70\\
& -ta=loadcache:L1 & -ta=loadcache:L1 & -ta=loadcache:L1 \\
& -acc -Mipa=fast &-acc -Mipa=fast &-acc -Mipa=fast \\
& -fastsse -O3 & -fastsse -O3& -fastsse -O3\\ 
 MPI & OpenMPI 1.10.7 & OpenMPI 1.10.7 & OpenMPI 1.10.7 \\
\hline 
\end{tabular}
\end{center}
\end{small}
\end{table}

\clearpage

\section{Application Example}
\label{sct5}
% 本章では東北沖地震における地震時滑り量の推定を行う.
In this section, we demonstrate the use of the developed solver by estimating the coseismic fault slip distribution in the 2011 Tohoku-oki earthquake. 
% このような滑り量の推定は地震発生メカニズムを考慮するうえで重要であり，地殻の形状近似は結果に大きな影響を与えることが既往の研究~\cite{eqdisttohoku1},~\cite{eqdisttohoku2}によって指摘されていることから，高分解能による高詳細なモデルを用いた地殻変動計算が必要となる．\\
This estimation is important for considering earthquake generation processes. Previous studies~\cite{eqdisttohoku1},~\cite{eqdisttohoku2} have shown that approximation in the geometry of the crust significantly changes the slip distribution. Thus, conducting crustal deformation analysis reflecting local geometry is required.

% まず地震時滑り量の推定手法について述べる．仮定される断層面を$n$個の単位断層に分け，単位断層変位を基底として，推定する断層変位を展開する．すなわち，推定される断層変位は次式のように表現できる．
First, we describe the method used to estimate the coseismic fault slip distribution following a previous study~\cite{agata2016robust}.
% These problem settings for inverse analysis is the same as those in the previous study~\cite{agata2016robust}.
The assumed fault plane is divided into $n$ small unit faults, and the fault slip is expanded using these unit faults as bases:
\begin{equation}
{\bf x}=\sum_{i=1}^{n}a_{i}\phi_{i},
\end{equation}
% ただし${\bf x}$は断層変位ベクトル，$a_i$ は$i$番目の単位断層変位の重ね合わせ係数，$\phi_i$は単位断層変位ベクトルである．
where ${\bf x}$ is the fault slip distribution vector, $a_i$ is the coefficient for the $i^{th}$ unit fault slip, and $\phi_i$ is the  distribution vector of the $i^{th}$ unit fault slip.
% 地表面上の$m$個の観測点での地殻変動データが利用可能であるとする．
We assume that observation data are available on the crustal surface at $m$ points, and 
% 地震時の地殻変動は線形な弾性変形とみなせることに注意すると，単位断層変位$\phi_i$に対する地表面上の観測点$j$での変位応答を表すグリーン関数 を計算することによって，断層変位推定は$a_i$を未知数とする以下の連立一次方程式に帰着される．
that the coseismic crustal deformation can be regarded as a linear elastic deformation.
Using Green's function $g_{ji}$ (i.e., surface response on observation point $j$ for unit fault slip $\phi_i$), yields the following estimation of the slip distribution:
\begin{equation}
\label{inversion}
\left(
\begin{array}{c}
{\bf G}\\
\alpha {\bf L}\\
\end{array}
\right)
{\bf a}=\left(
\begin{array}{c}
{\bf d}\\
{\bf 0}\\
\end{array}
\right),
\end{equation} 
%ここで,${\bf G}$は$g_{ji}$を成分とする$m\times n$次元行列，${\bf d}$は観測点$j$での地殻変動観測データ$d_j$ を成分とする$m$次元ベクトルである．${\bf L}$は滑り分布平滑化の拘束条件を表す行列で，${\bf G}$行列がill-poseであるために導入される．$\alpha$はL-curve法~\cite{hansen1992analysis}によって定まる重み係数である．\\
where ${\bf G}$ is an $m\times n$ matrix with components $g_{ji}$, and ${\bf d}$ is an $m$ dimensional vector of crustal deformation data on observation point $j$. ${\bf L}$ is a smoothing matrix introduced because ${\bf G}$ is generally ill-posed.
$\alpha$ is a weighting factor defined using the L-curve method~\cite{hansen1992analysis}. 
% 要約すると，すべり量分布を推定するにはグリーン関数の各成分である$g_{ji}$が必要となる．In summary, the components of the Green's function $g_{ji}$ are required to estimate the coseismic slip distribution.
% これは$n$個の単位断層変位に対する地表面応答をそれぞれ計算することによって得られる．
These Green's functions are obtained by computing surface responses against $n$ unit fault slips.
% 一般には$n=10^2-10^3$のオーダーであるため，$10^2$回以上の地殻変動計算となり，高分解の地殻データから生成される自由度$10^8$程度のモデルを使用する場合，計算コストは膨大なものとなる．
In typical problems, $n$ is of the order 10$^2$--10$^3$; thus, we must conduct crustal deformation computation more than 10$^2$ times. When we use finite-element models with 10$^8$ degrees of freedom required for reflecting the geometry of the crust, this simulation leads to huge computational cost.
% 一方で，この解析では同一の有限要素モデルに対して多数回の地殻変動計算を行うため，提案手法が適用可能である．
In this analysis, multiple crustal deformation computations are performed for the same finite-element model; thereby, use of the developed solver is expected to lead to high speedup.
% そこで，本章では提案手法による多数回地殻変動計算を高速に行い，提案手法の有用性を示す．\\
% Thus, in this section, we conduct multiple crustal deformation computation in a short time and show the effectiveness of our proposed method.

\begin{figure}[tb]
\begin{center}
 \includegraphics[clip,scale=0.75]{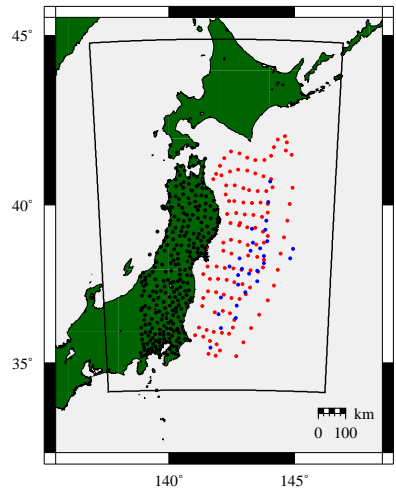}
     \caption{Target region of the application example (black line). The black, blue, and red points indicate the positions for GEONET, GPS-A, and S-net, respectively.}
     \label{appli_map}
\end{center}
\end{figure}

\begin{figure}[tb]
\begin{center}
 \includegraphics[clip,scale=0.65]{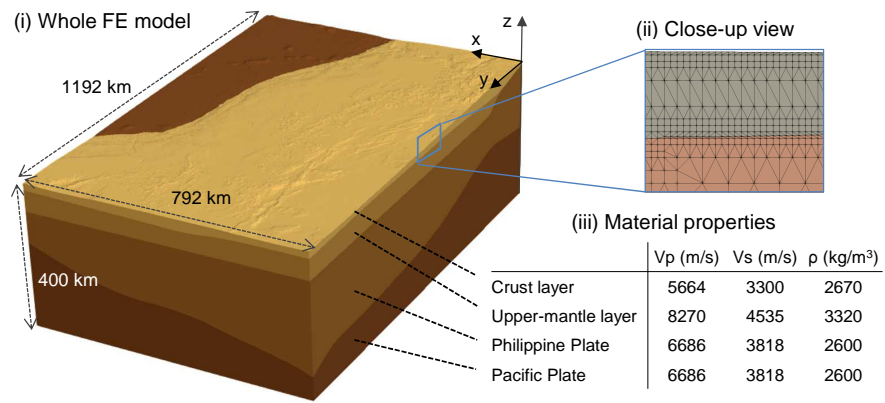}
     \caption{Finite-element model used for the application example.}% (XXX 図中のmateiralをmaterialに XXX)}
     \label{appli_model}
\end{center}
\end{figure}

\begin{figure}[tb]
 \begin{minipage}{0.47\hsize}
  \begin{center}
   \includegraphics[clip,scale=0.65]{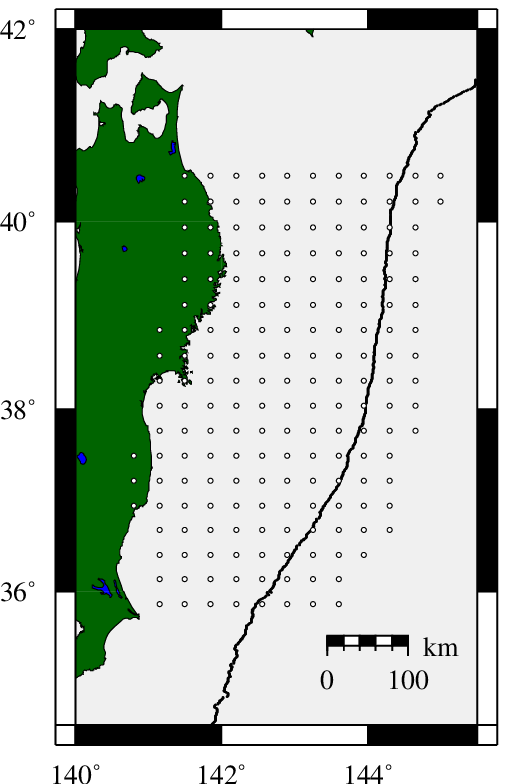}
  \end{center}
  \caption{Location of the centers of the unit fault slips.}
  \label{appli_faultloca}
 \end{minipage}
 \begin{minipage}{0.53\hsize}
   \begin{center}
   \includegraphics[clip,scale=0.55]{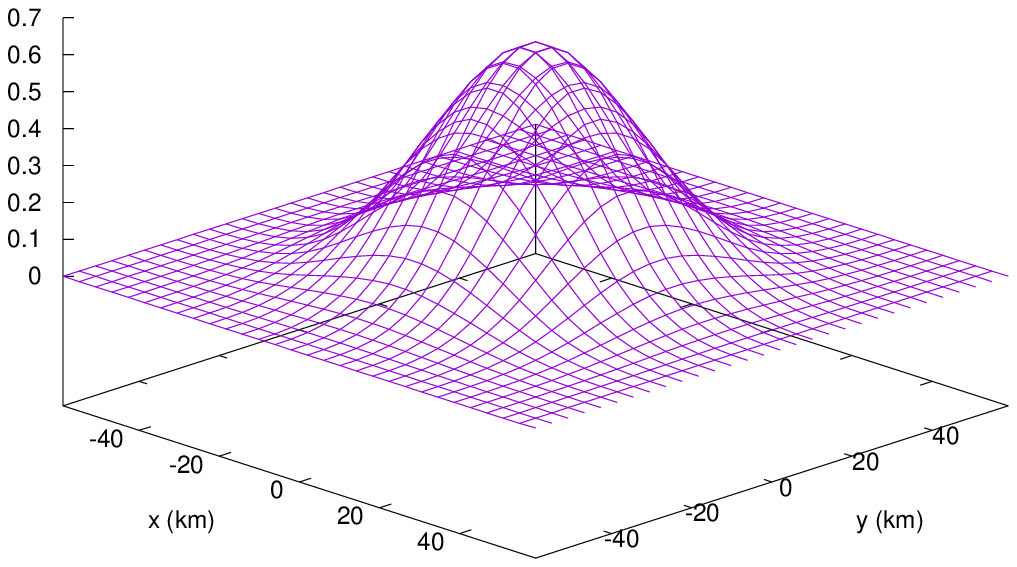}
  \end{center}
  \caption{Distribution of unit fault slip.}
  \label{appli_faultslip}
 \end{minipage}
\end{figure}

\begin{figure}[tb]
\begin{center}
 \includegraphics[clip,scale=0.75,angle=270]{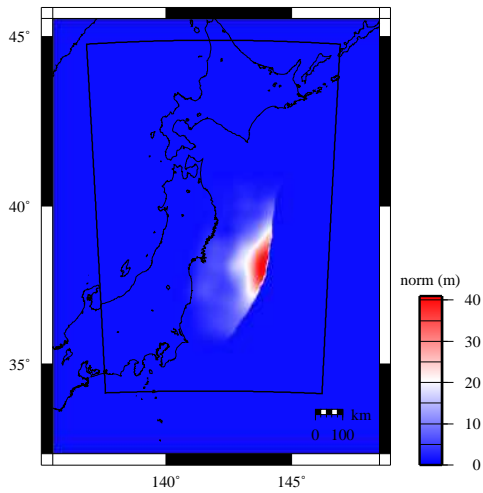}
     \caption{Estimated coseismic slip distribution.}
     \label{appli_res}
\end{center}
\end{figure}

% 対象とする領域は，図~\ref{appli_map}に示されており，792km×1192km×400kmの範囲，分解能が最小で1000 mとなる．
% また対象領域は4層からなる．
% 以上の条件から，図~\ref{appli_model}に示されるような有限要素モデルが生成される．
% モデルの自由度は409,649,580となり，四面体要素数100,494,786となった．
The four-layered 792 km $\times$ 1,192 km $\times$  400 km target area is shown in Fig.~\ref{appli_map}.
Modeling this area with a resolution of 1,000 m leads to a finite-element model consisting of 409,649,580
degrees of freedom and 100,494,786 tetrahedral elements (Fig.~\ref{appli_model}).
% 観測点にはGEONET，GPS-AおよびS-Netを利用し，それぞれx, y, z成分，x, y成分，z成分を用いる．
% We use GEONET, GPS-A, and S-Net as observation points. 
We use the $x$, $y$, and $z$ components of GEONET, the $x$ and $y$ components of GPS-A, and the $z$ component of S-net for the observed crust-deformation data.
% 単位すべりの入力地点は図~\ref{appli_faultloca}に表わされ，各点につきx, yの2方向に単位すべりを入力する．
The locations of the 184 input unit fault slips are shown in Fig.~\ref{appli_faultloca}.
For each point, Green's functions with unit B-spline function fault slips (Fig.~\ref{appli_faultslip}) are computed in the dip and strike directions.
% 単位滑りとしてはBスプライン関数を用いた図~\ref{appli_faultslip}のような滑りを入力する．
% これらの問題設定およびすべり量の推定手法は既往の研究~\cite{agata2016robust}と同様のものを使用している．
Thus, the total number of Green's functions becomes $n$ = 184 $\times$ 2 = 368.
% 適用例の計算については，前章のReedbush-Hの32ノードを用いて計算を行う．
We used 32 nodes of Reedbush-H and 
% 本問題では$n=368$であり，すべり量の推定には，単位すべりに対する368のグリーン関数を計算する必要がある．
%In this case $n$=368, and thus the estimation of fault slip requires 368 Green's functions for unit fault slips.
% 368回の地殻変動計算を16本ベクトル地殻変動計算×23セットとして行うことによって，グリーン関数がそれぞれ得られる．
obtained 368 Green's functions by conducting 23 sets of crustal deformation computations with 16 vectors.
% 計算されたグリーン関数と観測点のデータとを用いて式~\ref{inversion}を解くことによって，図~\ref{appli_res}のように地震時すべり量が推定された．\\
Figure~\ref{appli_res} shows the estimated slip distribution obtained by solving Eq.~(\ref{inversion}) using the computed Green's functions and observed data, a result consistent with previous studies~\cite{agata2016robust}.

% なお解析内の連立一次方程式求解時間は368本ベクトル全体で828秒となった．
The computation time for solving systems of linear equations was 828 s for 368 crustal deformation computations.
% これは小規模計算機環境を用いて行った同種の地殻変動計算~\cite{waccpd2016}と問題サイズ当たりで比較するとXXX倍の高速化となっている．
This computation is 29.2 times better in performance (5.11 times larger problem $\times$ solved 5.71 times faster) than previous studies on smaller computational environments in \cite{waccpd2016} with eight K40 GPUs, which conducted 360 crustal deformation computations with 80 million degrees of freedom in 4,731 s.
% 大きく計算時間が短縮しており，今回の適用例に加えて物性・形状に曖昧さを考慮したモンテカルロシミュレーションの適用や，地殻構造の最適化など，$10^6$回のオーダーでの地殻変動計算を必要とする解析についても，今後の実現可能性が十分あることを示せた．
From here we can see that the developed solver enabled reduction in computation time for a practical problem. In the future, we plan to use this method to optimize the crustal structure based on 10$^6$ cases of Monte Carlo crustal deformation computations with varying geometries and material properties.

\clearpage

\section{Concluding Remarks}
\label{sct6}

In this paper, we accelerated a low-order unstructured 3D finite-element solver targeting multiple large-scale crustal deformation analyses.
When we introduce accelerators, it is important to redesign the algorithm as its computer architecture greatly changes.
Based on a CPU-based solver attaining high performance on the K computer, we developed the solver algorithm more appropriate for a GPU architecture and then ported the code using OpenACC.
Here, we changed the algorithm such that multiple cases of finite-element simulations are conducted simultaneously thereby reducing random access and memory transfer per simulation case.
When the runtime on 20 K computer nodes and ten Reedbush-H nodes (20 P100 GPUs) were compared, the directly ported solver attained 5.0 times speedup, and the ported solver with modification to the algorithm attained 14.2 times speedup. We confirm that this modification is important to exhibit high performance in P100 GPUs and more effective for GPU-based Reedbush-H than for CPU-based K computer.
The developed solver is also highly effective on the Volta GPU architecture; we confirmed 2.52 times speedup with respect to eight P100 GPUs to eight V100 GPUs.
This acceleration enabled 368 crustal deformation computations targeting northeast Japan with 400 million degrees of freedom in 828 s on 32 Reedbush-H nodes, which is significantly faster than in the previous study.
The entire procedure of algorithm modification and OpenACC directive insertion was completed within two weeks; hence, we can see that high-performance gain can be attained with low development cost by using a suitable porting strategy.
Fast computations realized by the developed method are expected to be useful for quality assurance of earthquake simulations in the future.

\section*{Acknowledgments}
We thank Mr. Craig Toepfer (NVIDIA) and Mr. Yukihiko Hirano (NVIDIA) for the generous support and performance analyses concerning the use of NVIDIA DGX-1 (Volta V100 GPU) and NVIDIA DGX-1 (Pascal P100 GPU) environment. Part of the results were obtained using the K computer at the RIKEN Advanced Institute for Computational Science (Proposal numbers: hp160221, hp160160, 160157, and hp170249). This work was supported by Post K computer project (priority issue 3: Development of Integrated Simulation Systems for Hazard and Disaster Induced by Earthquake and Tsunami), Japan Society for the Promotion of Science (KAKENHI Grant Numbers 15K18110, 26249066, 25220908, and 17K14719) and FOCUS Establishing Supercomputing Center of Excellence.

\bibliographystyle{unsrt}
\bibliography{201708}

\end{document}